\documentclass{vietnam}

\usepackage{xspace}

\providecommand{\BLACKHAT} {{\textsc{BlackHat}}\xspace}
\providecommand{\HERWIG} {{\textsc{Herwig}}\xspace}
\providecommand{\JETPHOX} {{\textsc{JetPhox}}\xspace}
\providecommand{\MCATNLO} {{\textsc{MC@NLO}}\xspace}
\providecommand{\POWHEG} {{\textsc{POWHEG}}\xspace}
\providecommand{\PYTHIA} {{\textsc{Pythia}}\xspace}
\providecommand{\SHERPA} {{\textsc{Sherpa}}\xspace}
\providecommand{\TGNNLO} {{\textsc{2$\gamma$NNLO}}\xspace}
\providecommand{\HT}{\ensuremath{H_\mathrm{T}}\xspace}
\providecommand{\NRatio}{\ensuremath{N_\mathrm{32}}\xspace}
\providecommand{\RRatio}{\ensuremath{R_\mathrm{32}}\xspace}
\providecommand{\RdeltaR}{\ensuremath{R_{\Delta R}}\xspace}
\providecommand{\GeV}{\ensuremath{\,\mathrm{Ge\hspace{-.08em}V}}\xspace}
\providecommand{\GeVsq}{\ensuremath{\,\mathrm{Ge\hspace{-.08em}V}^2}\xspace}
\providecommand{\TeV}{\ensuremath{\,\mathrm{Te\hspace{-.08em}V}}\xspace}

\providecommand{\alpsq}{\ensuremath{\alpha_S(Q)}\xspace}
\providecommand{\alpsmz}{\ensuremath{\alpha_S(M_Z)}\xspace}
\providecommand{\avept}{\ensuremath{\langle p_\mathrm{T1,2}\rangle}\xspace}

\providecommand{\kt}{\ensuremath{k_\mathrm{T}}\xspace}
\providecommand{\pt}{\ensuremath{p_\mathrm{T}}\xspace}

\newcommand{\Photo}{\includegraphics[height=35mm]{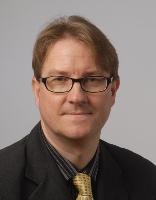}}

\begin{document}
\vspace*{4cm}\title{EXPERIMENTAL TESTS OF QCD}

\author{K. RABBERTZ\\(on behalf of the ATLAS and CMS Collaborations)}

\address{Karlsruher Institut f{\"u}r Technologie,
  Institut f{\"u}r Experimentelle Kernphysik,\\
  Campus S{\"u}d, Postfach 6980, D-76128 Karlsruhe, Germany}

\maketitle\abstracts{%
  The first very successful LHC running period has been finished. At
  $7\,$TeV centre-of-mass energy about $5\,\mathrm{fb}^{-1}$ of data
  have been collected and at $8\,$TeV even
  $20\,\mathrm{fb}^{-1}$. Many detailed analyses of
  these data are still going on. The latest measurements on photon,
  weak boson plus jet, and jet production are compared against the
  most recent theory predictions. They are complemented by new results
  reported by the experiments at the Tevatron and HERA
  colliders. Finally, several new determinations of the strong
  coupling constant from jet data are presented.%
}

\section{Introduction}
\label{sec:intro}

Through the abundant production of jets, i.e.\ collimated streams of
particles, hadron colliders essentially become \emph{jet
  laboratories}. QCD analyses of these data comprise a huge variety of
phenomena and allow, amongst others, to learn about hard QCD or
nonperturbative effects. As background the huge cross sections of jet,
photon, and weak boson production pose a problem for many searches for
new phenomena. Since hadrons are ``made of QCD'', a precise
understanding of their structure is indispensable as a linking piece
between many collision types and processes. In particular at the LHC,
which makes a huge new region of phase space accessible in energy
scale $Q$ and fractional momentum of the proton $x$, new measurements
by ATLAS and CMS~\cite{Aad:2008zzm,Chatrchyan:2008aa} provide
significant constraints on the parton distribution functions (PDFs) of
the proton and the strong coupling constant \alpsmz.

\section{Photon production}
\label{sec:photons}

The analysis of direct photons is complicated by the fact that photons
can also be produced in the transition phase from a parton shower to
measurable particles.
This so-called fragmentation component is much less understood
than the perturbatively calculable QCD Compton, $qg \rightarrow \gamma
q$, or annihilation processes, $q\bar{q} \rightarrow \gamma g$.
Despite the possibility to probe and better determine the gluon PDF,
observed discrepancies between measurement and theory for
centre-of-mass energies of 20--40\GeV lead to the abandonment of
photon data for PDF analyses.\cite{d'Enterria:2012yj} With improved
techniques requiring photons to be isolated, the fragmentation
component can be suppressed, in particular when accessing much higher
photon transverse momenta at the LHC with more fine-grained detectors
than previously possible. This suggests to re-address the comparison
of photon data to perturbative QCD and their re-integration into
global PDF fits.

\begin{figure}[tb]
  \centering
  \includegraphics[width=0.49\linewidth]{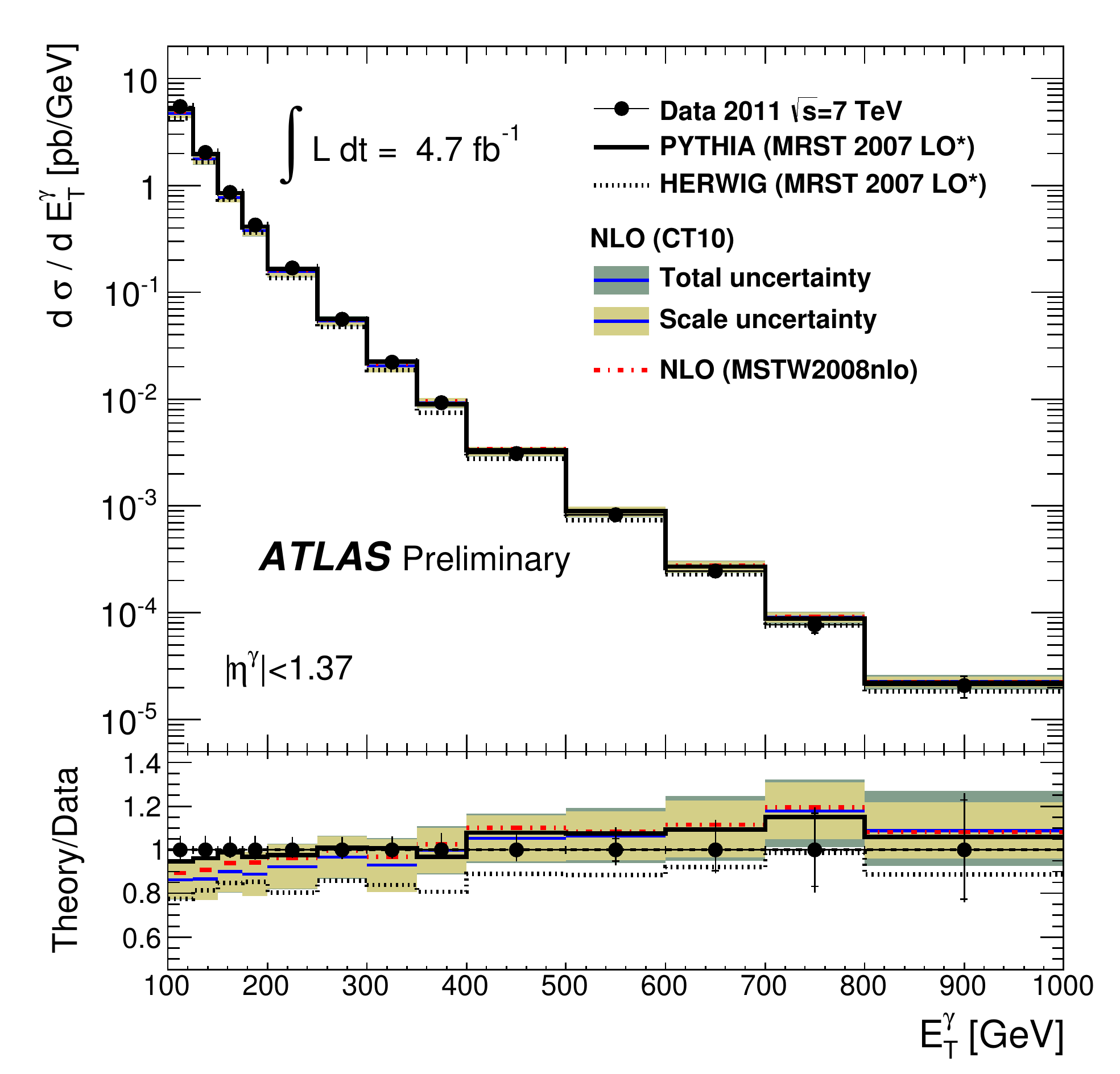}%
  \includegraphics[width=0.49\linewidth]{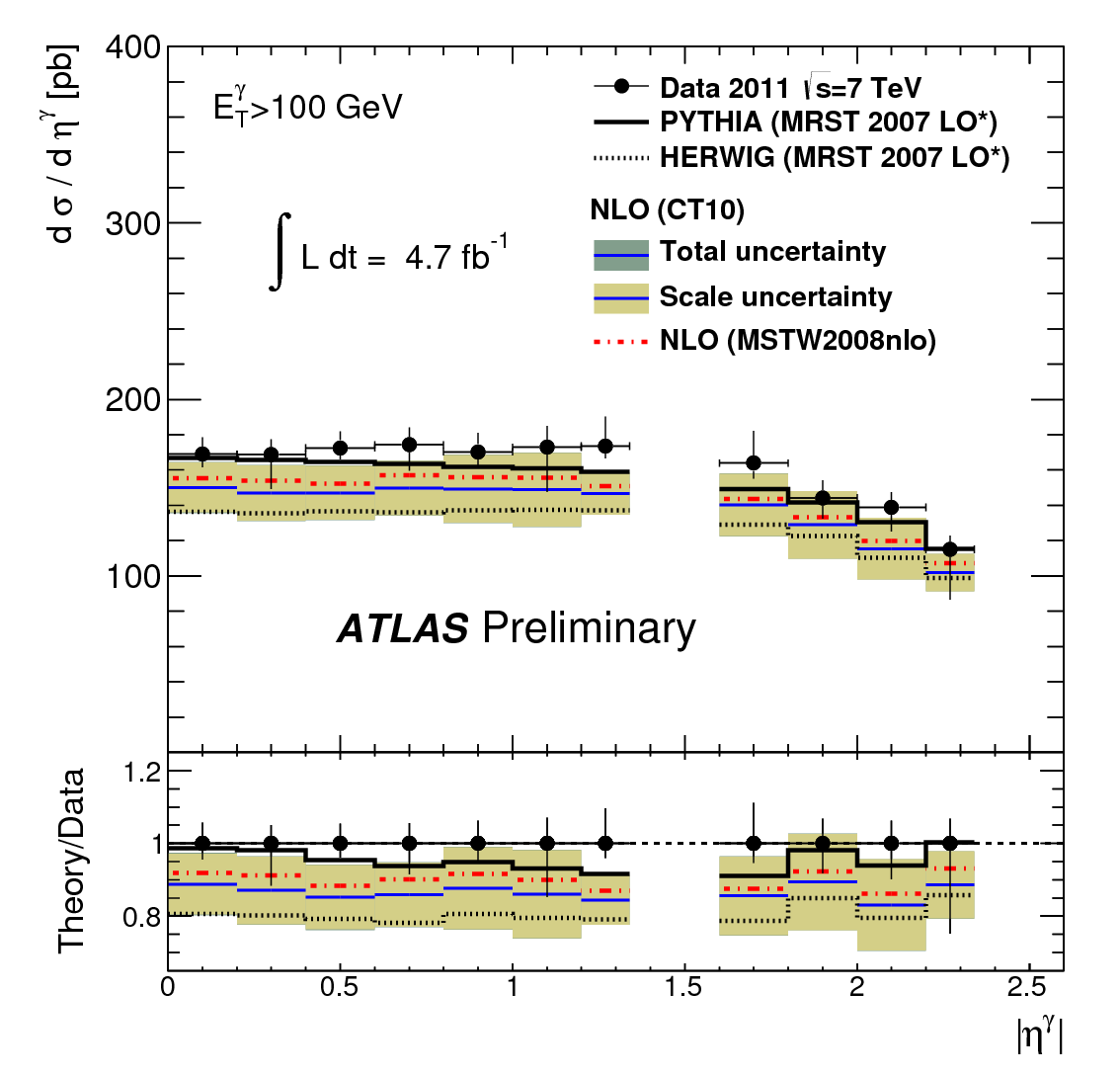}
  \caption{Transverse energy and absolute rapidity of isolated prompt
    photons as measured by ATLAS~\protect\cite{ATLAS-CONF-2013-022} in
    comparison to LO predictions from
    \PYTHIA~\protect\cite{Sjostrand:2006za} and
    \HERWIG,\protect\cite{Marchesini:1991ch} and to NLO from
    \JETPHOX.\protect\cite{Aurenche:2006vj}}
  \label{fig:pgamma}
\end{figure}

In Fig.~\ref{fig:pgamma} left the ATLAS Collaboration extends the
measured cross section of isolated photons versus transverse energy up
to 1\TeV and observes agreement with NLO predictions by
\JETPHOX~\cite{Aurenche:2006vj} over five orders of magnitude in cross
section. Some tension between data and theory remains with respect to
the absolute photon rapidity as shown in Fig.~\ref{fig:pgamma}
right. The limiting factor for more precise comparisons lies with the
dominant scale uncertainty of the NLO calculations.

Within the context of Higgs boson searches (and now measurements) the
isolated photon-pair production as irreducible background to $H
\rightarrow \gamma\gamma$ is of special interest. The CDF and D0
Collaborations have published detailed studies including the full
recorded luminosity~\cite{Aaltonen:2012jd,Abazov:2013pua} and find a
significantly improved description at NNLO from
\TGNNLO~\cite{Catani:2011qz} compared to NLO\@.

\begin{figure}[h!]
  \centering
  \includegraphics[width=0.43\linewidth]{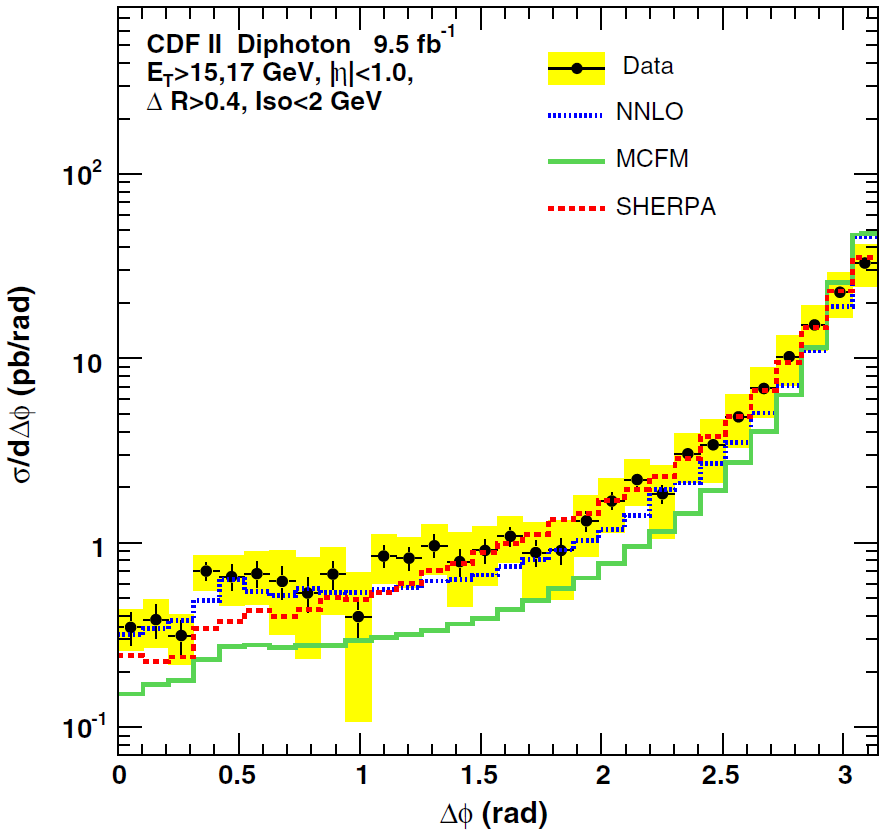}%
  \includegraphics[width=0.56\linewidth]{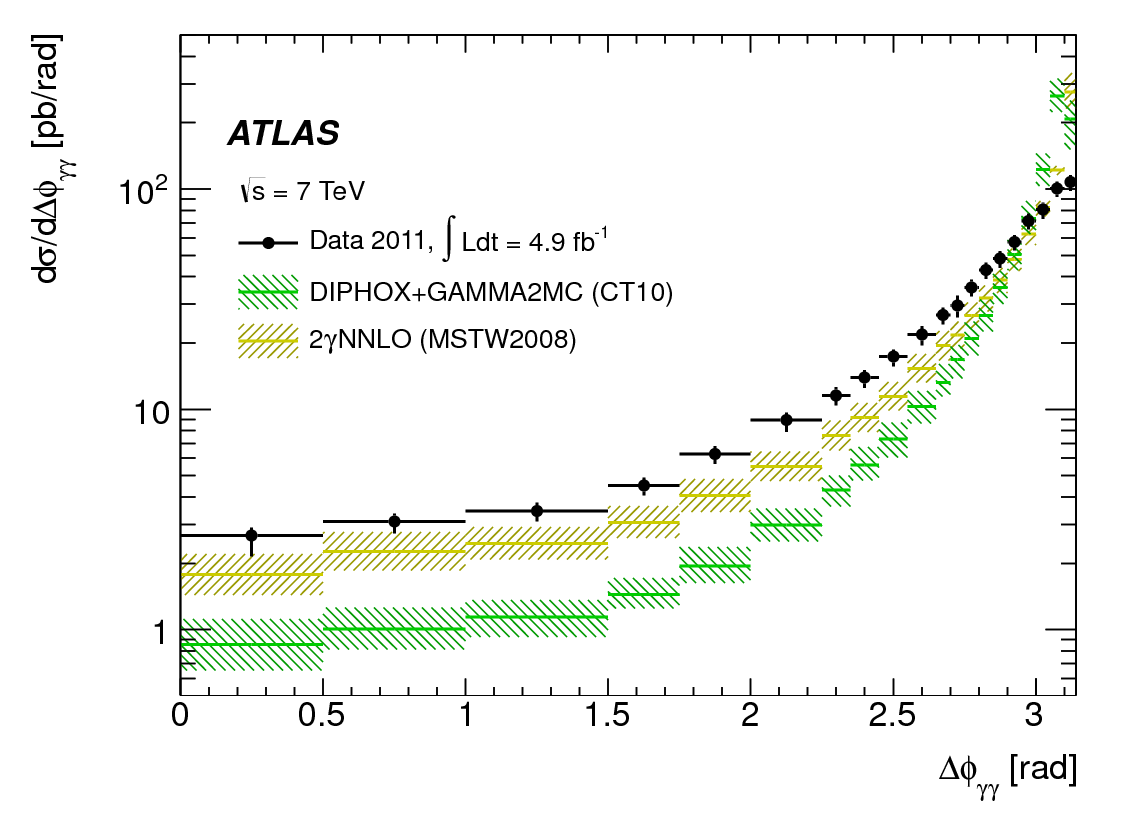}
  \caption{Azimuthal angular difference between the photons of
    isolated photon-pair events. The differential cross sections,
    measured by CDF at 1.96\TeV (left)~\protect\cite{Aaltonen:2012jd}
    and by ATLAS at 7\TeV (right)~\protect\cite{Aad:2012tba} are
    compared to predictions by theory up to NNLO\@.}
  \label{fig:digamma}
\end{figure}

This is confirmed by ATLAS,\cite{Aad:2012tba} although some
discrepancies remain in particular in the distribution of the
azimuthal difference between the two photons as presented in
Fig.~\ref{fig:digamma} for CDF (left) and ATLAS (right).

\section{Weak boson+jet production}
\label{sec:bosonjet}

\begin{figure}[tbp]
  \centering
  \includegraphics[width=0.51\linewidth]{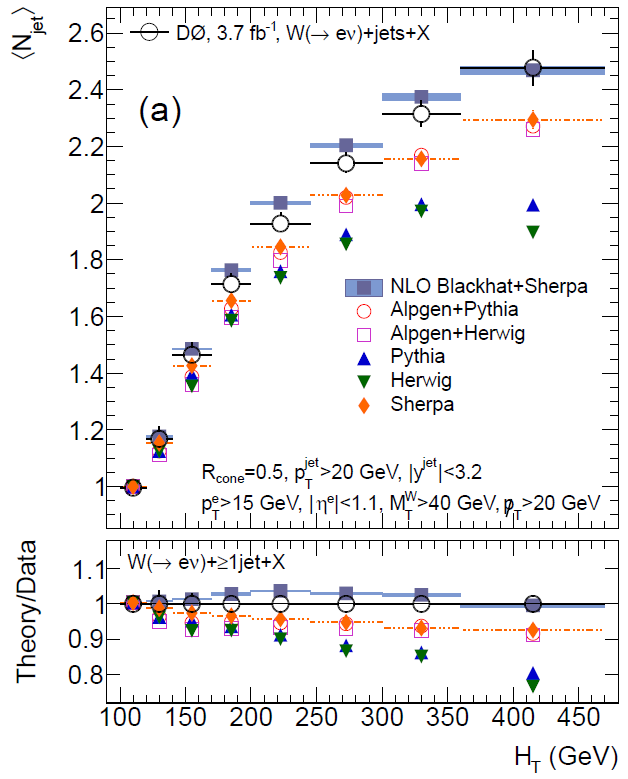}%
  \includegraphics[width=0.48\linewidth]{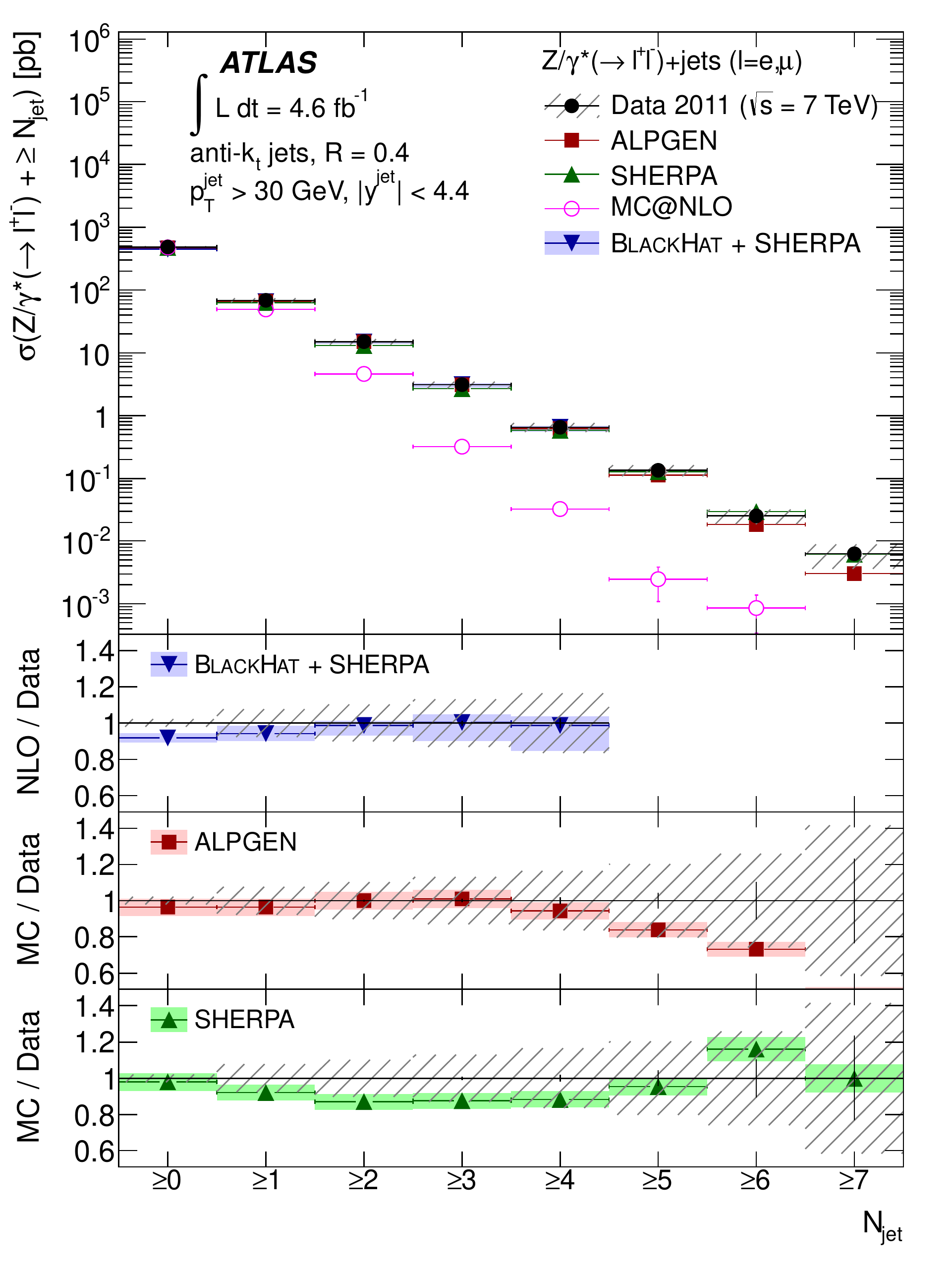}
  \caption{Average jet multiplicity versus scalar jet \pt sum \HT in
    $W$ events from D0 at 1.96\TeV~\protect\cite{Abazov:2013gpa} and
    inclusive production cross section of a $Z$ boson together with
    zero or more up to seven or more jets from ATLAS at
    7\TeV.\protect\cite{Aad:2013ysa} The measurements are compared to
    various predictions including \BLACKHAT +
    \SHERPA,\protect\cite{Berger:2010vm,Berger:2010zx,Ita:2011wn,Gleisberg:2008ta}
    which provides NLO precision up to inclusive $W,Z$ plus four jet
    production.}
  \label{fig:vjets}
\end{figure}

\begin{figure}[tbp]
  \centering
  \includegraphics[width=0.42\linewidth]{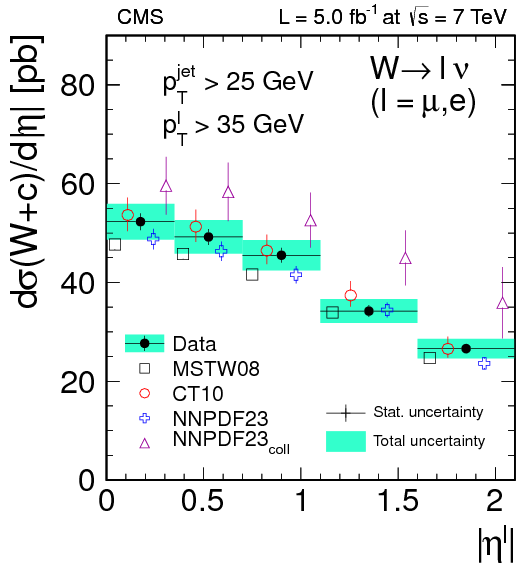}%
  \includegraphics[width=0.57\linewidth]{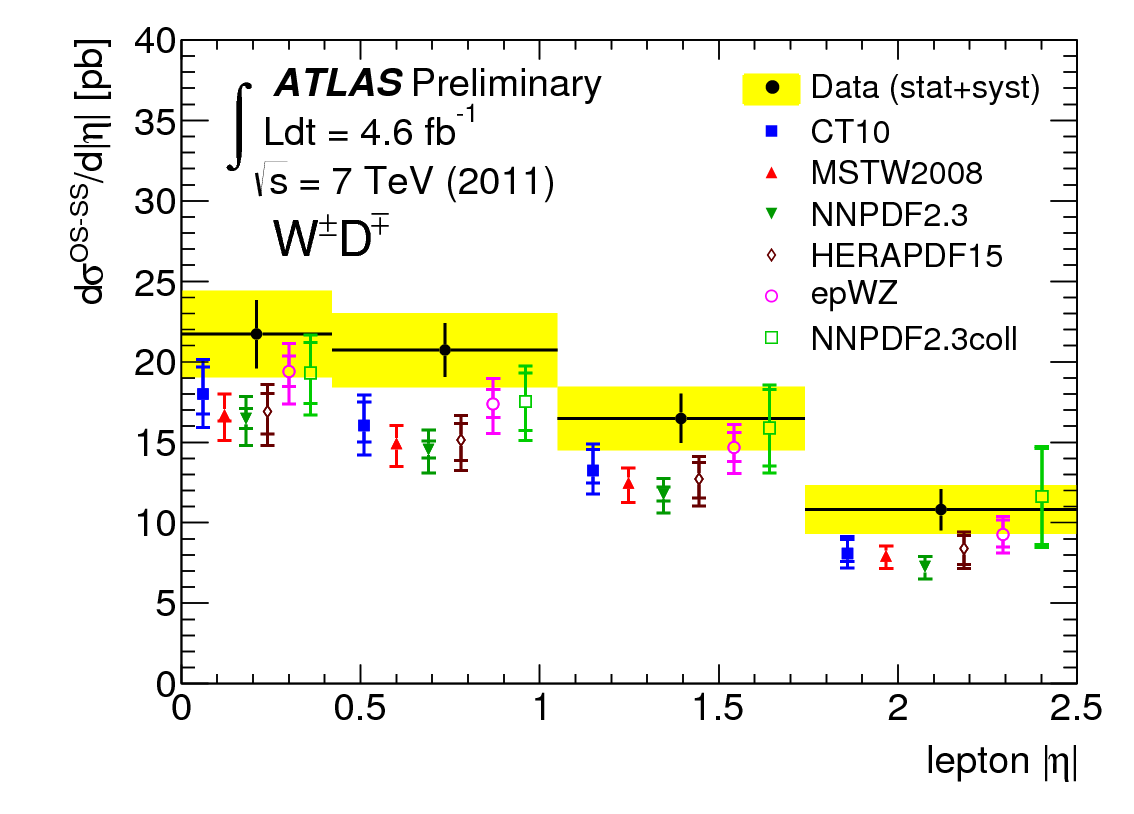}
  \caption{Comparison of associated $W$ plus charm production cross
    sections, differential in absolute lepton rapidity, from CMS
    (left)~\protect\cite{Chatrchyan:2013uja} and ATLAS
    (right)~\protect\cite{ATLAS-CONF-2013-045} versus theory
    predictions using diverse PDF sets. The NNPDF2.3$_\mathrm{coll}$
    set does not include constraints on the strange PDF from
    low-energy DIS data.}
  \label{fig:wplusc}
\end{figure}

The D0 Collaboration also investigated the production of $W$ bosons in
association with at least one jet,\cite{Abazov:2013gpa} which is well
described by NLO as given by
\BLACKHAT + \SHERPA.\cite{Berger:2010vm,Berger:2010zx,Gleisberg:2008ta}
As expected LO plus parton shower event generators fail progressively
for higher jet multiplicities as demonstrated in Fig.~\ref{fig:vjets}
left. The ATLAS experiment measured inclusive $Z+$jet cross sections
with up to seven or more jets.\cite{Aad:2013ysa} Again \BLACKHAT +
\SHERPA~\cite{Berger:2010vm,Ita:2011wn,Gleisberg:2008ta} provides NLO
precision up to inclusive $Z+4$-jet production and agrees with the
data, while \MCATNLO~\cite{Frixione:2002ik} exhibits large
discrepancies for more than one jet, see Fig.~\ref{fig:vjets} right.

Identifying charm jets produced in association with a $W$ boson, e.g.\
via decays of the charmed hadrons $D^\pm$ or $D^{*\pm}$, gives direct
access to the strange quark and antiquark content of the proton. This
can help to significantly reduce the uncertainties of and assess
potential asymmetries between the strange quark and antiquark PDFs\@.
The measurement by CMS~\cite{Chatrchyan:2013uja} as presented in
Fig.~\ref{fig:wplusc} left agrees with theory computations at NLO from
MCFM~\cite{Campbell:2010ff} employing global PDF sets that include
constraints on the strange content from low-energy deep-inelastic
scattering (DIS) data. In contrast, a similar study from
ATLAS~\cite{ATLAS-CONF-2013-045} shown in Fig.~\ref{fig:wplusc} right
prefers PDFs excluding these DIS data. Since the phase space of the
two analyses is different, their compatibility is hard to judge. It
will be illuminating to see both data sets, when finalized, included
into a common global PDF fit.

\section{Jet production}
\label{sec:jets}

\begin{figure}[tbp]
  \centering
  \includegraphics[width=0.50\linewidth]{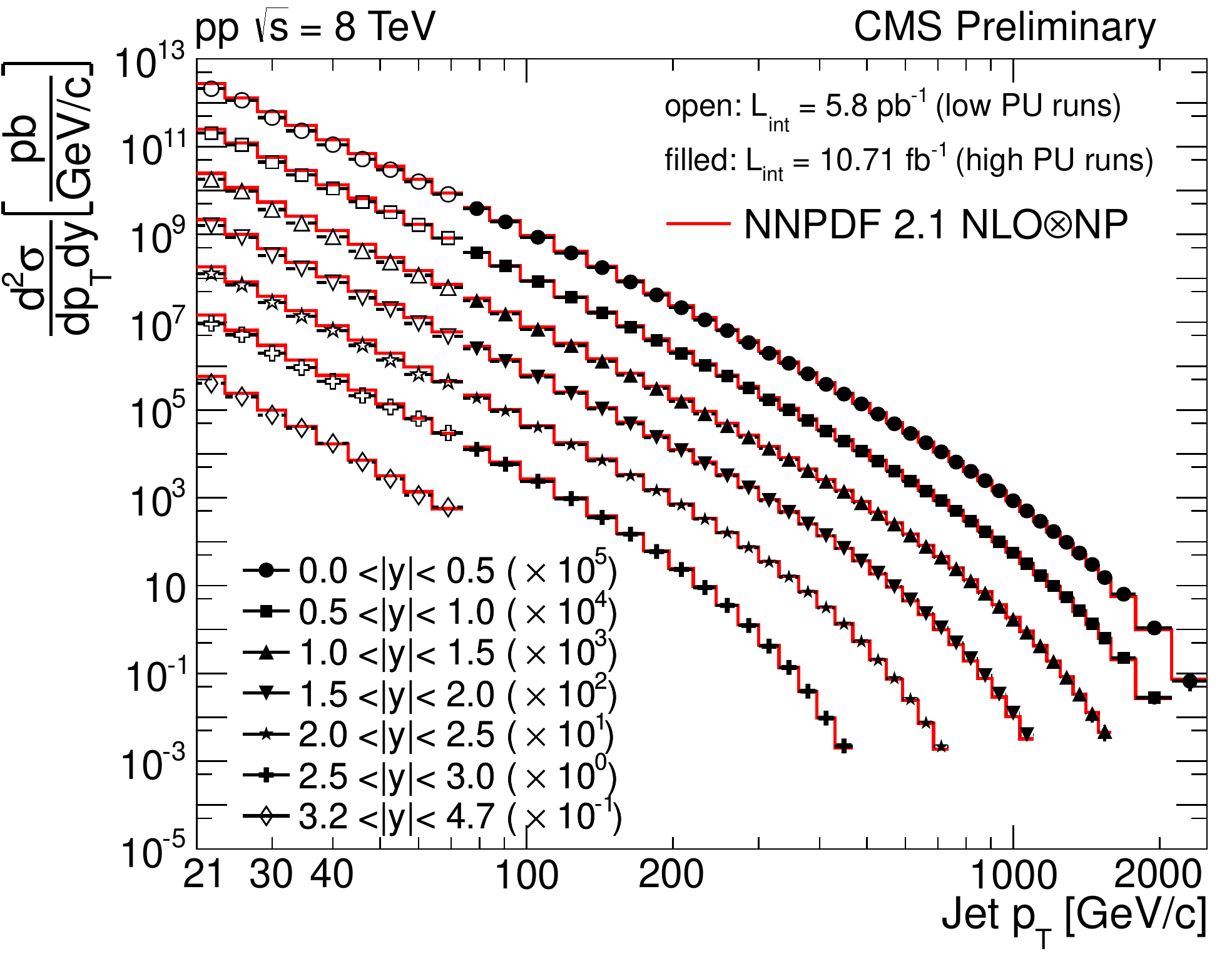}\hfill%
  \includegraphics[width=0.48\linewidth]{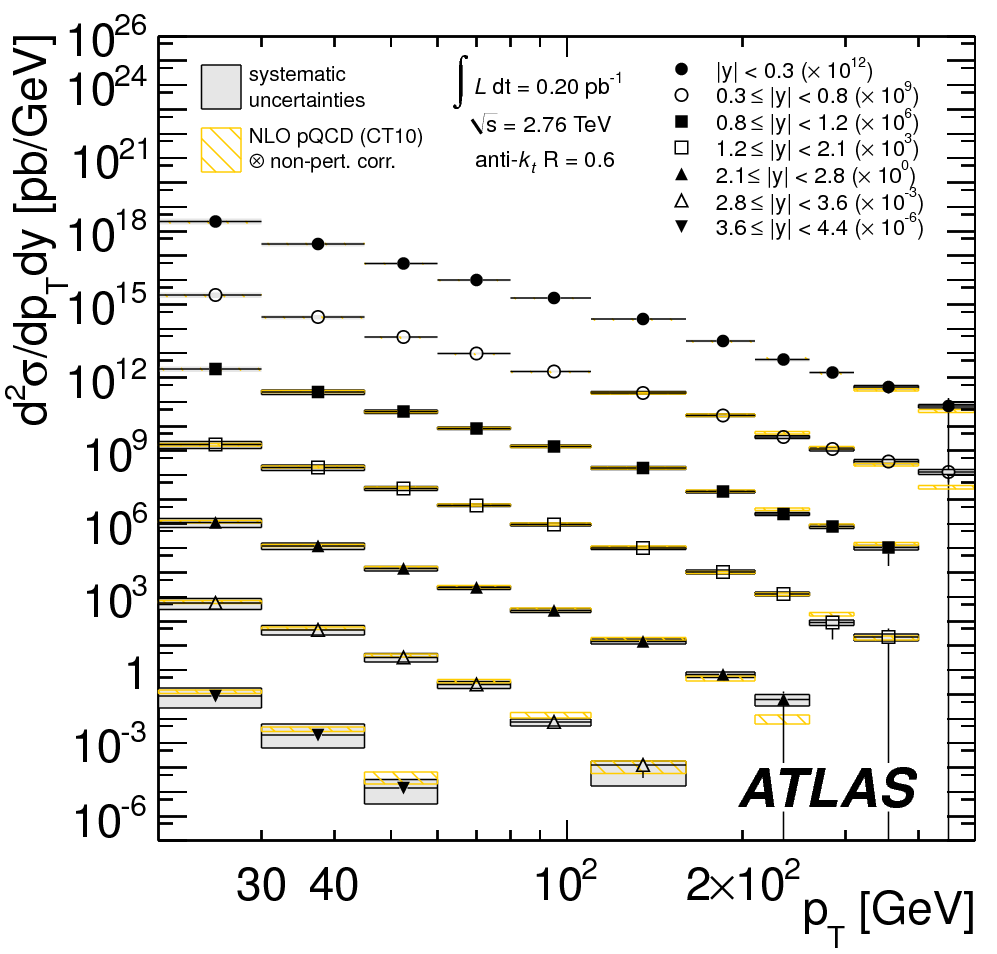}
  \caption{Double-differential inclusive jet cross sections as
    measured by
    CMS~\protect\cite{CMS-PAS-SMP-12-012,CMS-PAS-FSQ-12-031} at 8\TeV
    and by ATLAS~\protect\cite{Aad:2013lpa} at 2.76\TeV. The
    measurements are compared to predictions at NLO times
    nonperturbative (NP) corrections.}
  \label{fig:inclusivejets}
\end{figure}

\begin{figure}[tbp]
  \centering
  \includegraphics[width=0.49\linewidth]{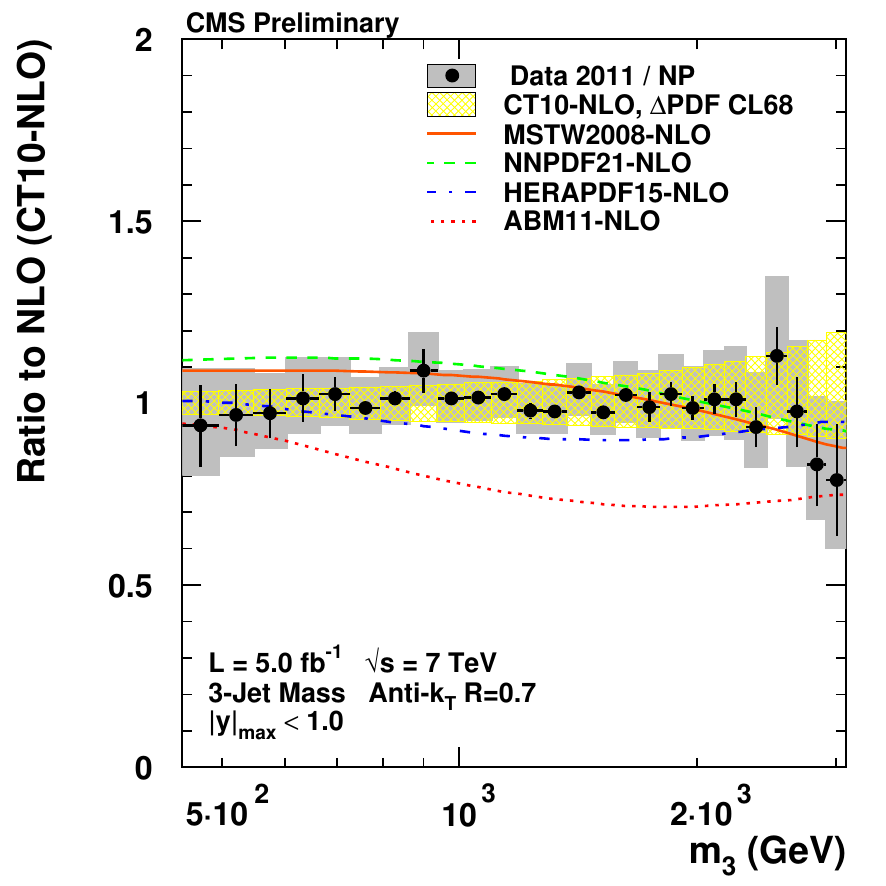}%
  \includegraphics[width=0.49\linewidth]{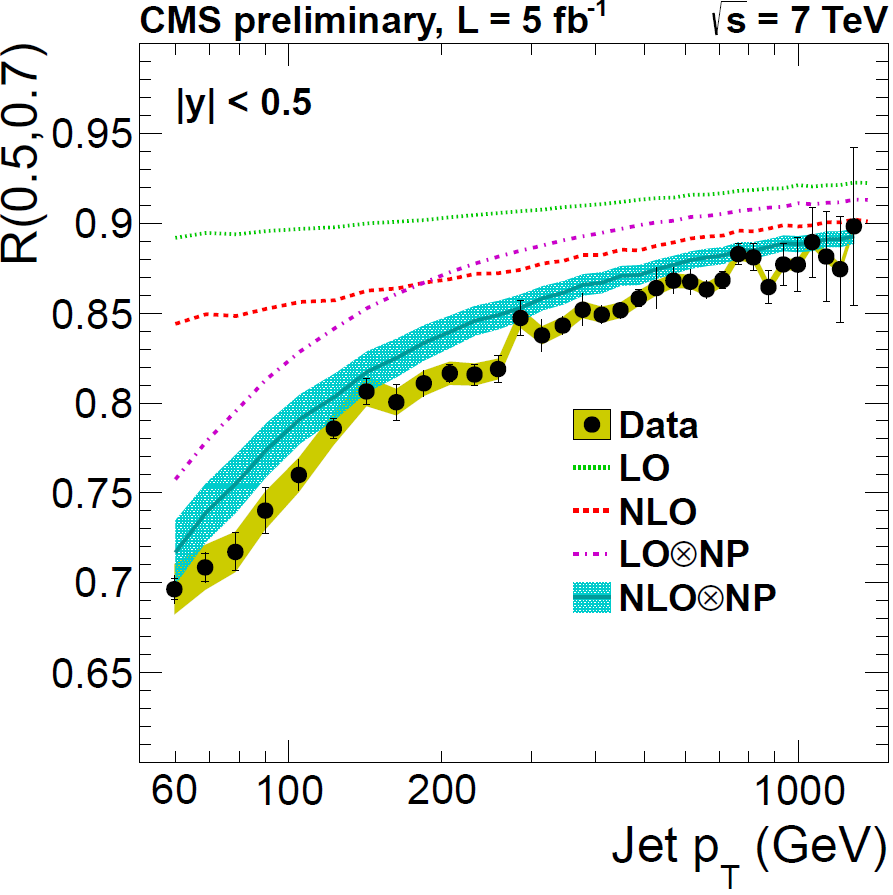}
  \caption{The 3-jet mass cross section for central rapidity by CMS
    (left)~\protect\cite{CMS-PAS-SMP-12-027} at 7\TeV as a ratio to
    theory at NLO times nonperturbative (NP) corections for various
    PDF sets, and the jet cross-section ratio, also at 7\TeV, for the
    two different jet size parameters of $R=0.5$ and $0.7$ by CMS
    (right)~\protect\cite{CMS-PAS-SMP-13-002}. The latter is compared
    to predictions at LO and NLO with or without NP corrections; the
    best description of the data is given by
    \POWHEG\protect\cite{Alioli:2010xa} NLO + \PYTHIA (not shown).}
  \label{fig:3-jetmassrratio}
\end{figure}

ATLAS and CMS, both employ the anti-\kt\ jet
algorithm~\cite{Cacciari:2008gp} to define their jets, however with
different jet size parameters $R$ of $0.4$ or $0.6$ for ATLAS and
$0.5$ or $0.7$ for CMS, respectively. Profiting from the excellent
performance of both detectors the dominant experimental uncertainty
induced by the jet energy calibration could be limited to about
10--20\%. The common normalization uncertainty caused by the
luminosity determination could be reduced from initially more than
10\% down to 2--4\%.  Contrasting jet measurements from
CMS,\cite{Chatrchyan:2012bja,CMS-PAS-SMP-12-012,CMS-PAS-FSQ-12-031}
e.g.\ at 8\TeV in Fig.~\ref{fig:inclusivejets} left, or from
ATLAS~\cite{Aad:2011fc,ATLAS-CONF-2012-021,Aad:2013lpa} at 2.76\TeV in
Fig.~\ref{fig:inclusivejets} right, with theory at NLO demonstrates
agreement within uncertainties. This statement holds also when looking
into higher order observables like the 3-jet mass cross section as
studied by CMS~\cite{CMS-PAS-SMP-12-027} in
Fig.~\ref{fig:3-jetmassrratio} left for predictions involving various
PDF sets. The limiting factor, however, for even more accurate
comparisons is the lack of NNLO predictions, which, at least for dijet
production, will become available in the near
future~\cite{Ridder:2013mf} and allows to reduce the large scale
dependence.

\begin{figure}[tbp]
  \centering
  \includegraphics[width=0.75\linewidth]{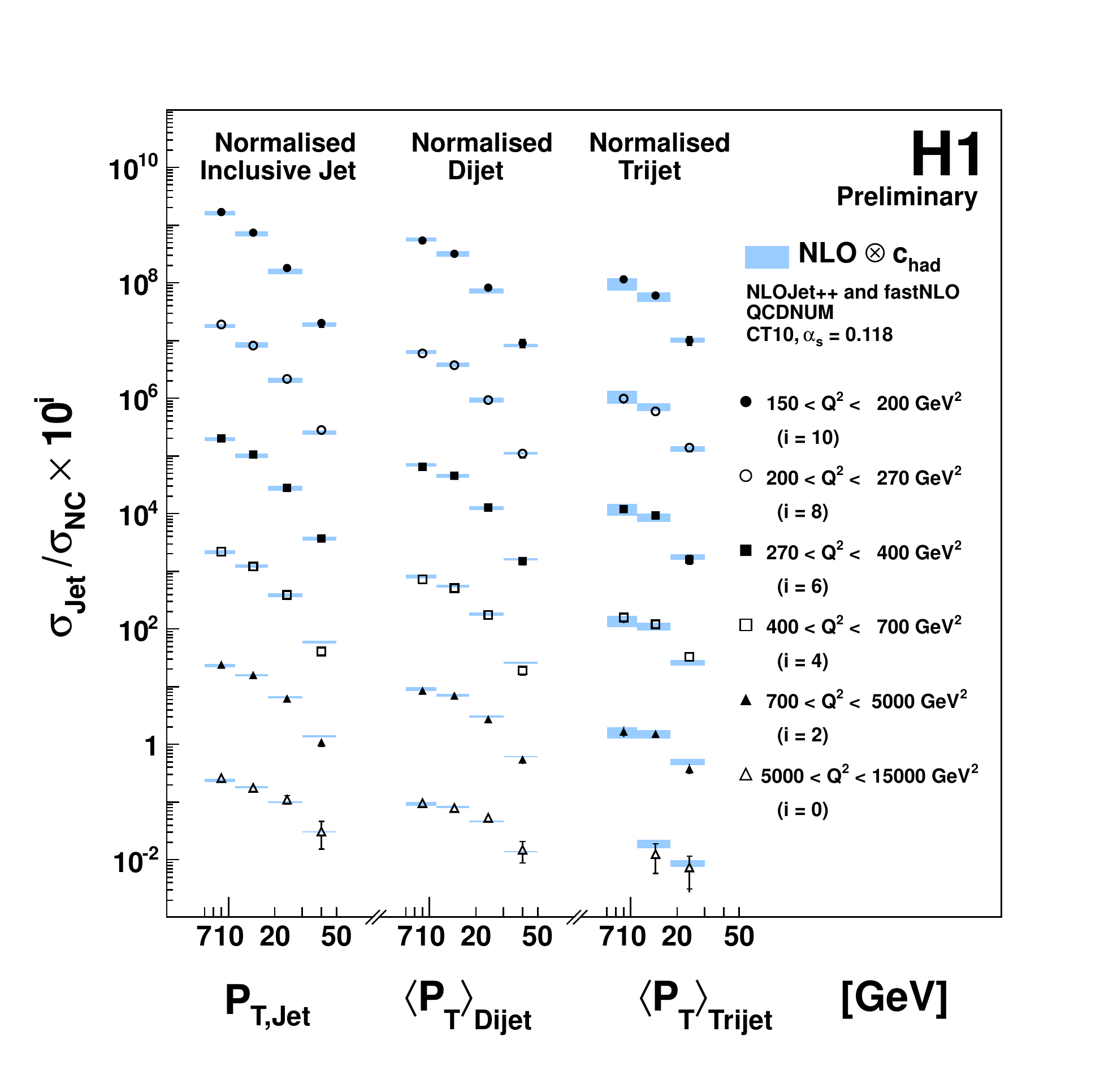}
  \caption{Inclusive jet, dijet, and trijet cross sections as measured
    by H1~\protect\cite{H1prelim-12-031} for momentum transfers
    squared $Q^2$ between 150 and 15000\GeVsq. The cross sections
    are normalized to the neutral-current cross section in
    deep-inelastic $ep$ scattering.}
  \label{fig:multijetsnorm}
\end{figure}

\begin{figure}[tbp]
  \centering
  \includegraphics[width=0.75\linewidth]{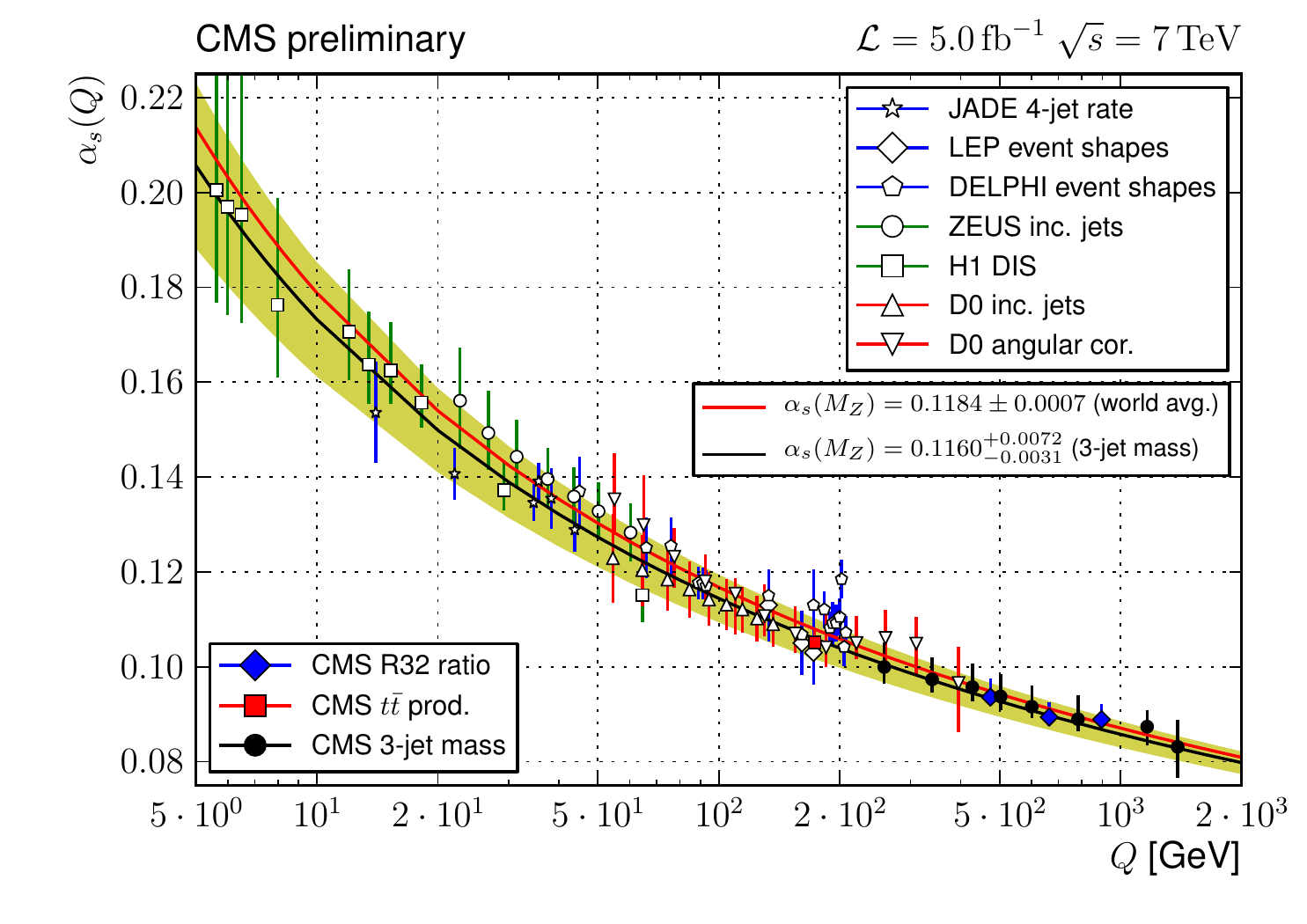}
  \caption{The strong coupling \alpsq\ (solid black line) and its
    total uncertainty (band) evolved from the CMS determination
    $\alpsmz=0.1160_{-0.0031}^{+0.0072}$ as a function of the momentum
    transfer $Q=\avept$. The extractions of \alpsq\ from the 3-jet
    mass measurement are shown in eight separate ranges of $Q$
    together with previous results from
    CMS~\protect\cite{Chatrchyan:2013txa,Chatrchyan:2013haa} and from
    other hadron collider
    experiments.\protect\cite{Aaron:2009vs,Aaron:2010ac,Abramowicz:2012jz,Abazov:2009nc,Abazov:2012lua}}
  \label{fig:asrunning}
\end{figure}

A possibility to partially cancel experimental as well as theoretical
uncertainties in cross-section ratios is exploited by ATLAS through
the division of their inclusive jet measurements at two different
energy points.\cite{Aad:2013lpa} Profiting from data at 2.76\TeV, the
baseline proton-proton centre-of-mass energy for heavy ion collisions,
in addition to the 7\TeV data, the ATLAS Collaboration derives more
significant constraints on PDFs in the accessible phase space than
when considering each jet cross section alone.

Another possibility is, following a suggestion in
Ref.,\cite{Soyez:2011np} the cross-section ratio for jets defined with
different jet size parameters. Using this method, details of the
parton showering and the nonperturbative hadronization phase are
emphasized such that even NLO calculations are not able to describe
the data as shown by CMS~\cite{CMS-PAS-SMP-13-002} in
Fig.~\ref{fig:3-jetmassrratio} right, or previously by the ALICE
experiment.\cite{Abelev:2013fn} Investigating the angular correlation
in the emission of third jets around the second-leading jet in \pt,
effects of colour coherence of the strong interaction can be compared
to the modelling in parton shower event generators. Such an analysis
has been performed recently by CMS~\cite{Chatrchyan:2013tmp}
emphasizing the need for better event generator tunings.

In contrast, observables like the multi-jet cross sections normalized
to the neutral-current DIS cross section~\cite{H1prelim-12-031} shown
in Fig.~\ref{fig:multijetsnorm}, the average number of neighbouring
jets within a given distance in an inclusive jet sample,
\RdeltaR,\cite{Abazov:2012lua} or the inclusive 3-jet over 2-jet
event, \RRatio,\cite{Chatrchyan:2013txa} or jet cross-section ratios,
\NRatio,\cite{ATLAS-CONF-2013-041} are designed to be reliably
comparable to perturbative QCD\@. Confronting theory predictions with
measurements of these quantities the strong coupling constant \alpsmz
can be determined from energy scales $Q$ ranging from tenths of GeV up
to 1\TeV\@. All fit results, $\alpsmz = 0.1163 ^{+0.0048}_{-0.0040}$ (H1),
$\alpsmz = 0.1191 ^{+0.0048}_{-0.0071}$ (D0), $\alpsmz = 0.111
^{+0.017}_{-0.007}$ (ATLAS), and $\alpsmz = 0.1148 \pm 0.0055$ (CMS),
where the quadratically added total uncertainty is given, are
compatible with the world average value of
$\alpsmz=0.1184\pm0.0007$.\cite{Beringer:1900zz} In all cases the
total uncertainty is dominated by the scale uncertainties of the
theory predictions at NLO\@.  An overview of the $Q$ dependent
determinations including the D0 and CMS results, the latter also
comprising \alpsmz extractions from 3-jet mass cross sections and
$t\bar{t}$ production at
threshold,\cite{CMS-PAS-SMP-12-027,Chatrchyan:2013haa} is shown in
Fig.~\ref{fig:asrunning}. No deviation from the expected running
behaviour of the strong coupling \alpsq is observed.

\section{Summary}
\label{sec:summary}

It is a great achievement that such a large amount of accurate data
from multiple colliders are in general agreement with predictions of
QCD over many orders of magnitude in cross section and in a wide
region of phase space. Although a vast number of new theoretical tools
and calculations, including multiplicities with up to five jets, have
been developed at the same time, there are still cases, where
theoretical uncertainties are the limiting factor. For even more
precise comparisons NNLO is required.
With all this progress in measurements and theory more insight can be
gained into the workings of QCD with significant impact on the strong
coupling constant, PDFs, and cross-section predictions e.g.\ for the
Higgs boson and for searches for new physics.

\section*{References}
\bibliography{vietnam}

\end{document}